\documentstyle[aps,prd]{revtex} 
\begin{document}

\def\oh{\Omega h^2 }
\def\ohsl{\oh_{\rm SLC}}
\def\ohds{\oh_{\rm NCC}}
\def\ohall{\oh_{\rm BCC}}
\def\ohno{\oh_{\rm IGC}}
\def\ohco{\oh_{\rm COA}} 
\def\mstau{m_{\t \tau}}
\def\mchar{m_{\chi^+}}
\def\seff{\langle \sigma_{\rm eff} v \rangle}
\def\sann{\langle \sigma_{\chi\chi} v \rangle} 
\def\sigmaav{\langle \sigma_{\chi\chi} v \rangle}
\def\sigmaijav{\langle \sigma_{ij}v_{ij} \rangle}
\def\sigmagen{\langle \sigma v \rangle}
\def\mchar{m_{\t\chi^\pm}}
\def\mchi{m_{\chi} }
\def\gev{\mbox{ GeV}} 
\def\tev{\mbox{ TeV}}
\def\dtgev{\,\mbox{GeV}^{-2}}
\def\dgev{{1\over \mbox{m}^2\cdot \mbox{yr}}}
\def\hgev{\,\mbox{m}^{-2}\cdot \mbox{yr}^{-1}} 
\def\events{\, {\rm event}/({\rm kg}\cdot {\rm day})}
\def\pur{{\cal P}}
\def\t{\tilde}
\def\eq{{\scriptsize{\mbox{eq}}}}
\def\etal{{\it {et\ al.}}} 
\def\WT{\widetilde} 
\title{Slepton and Neutralino/Chargino Coannihilations in MSSM}

\author{V.A.~Bednyakov}
\address{Laboratory of Nuclear Problems,
         Joint Institute for Nuclear Research, \\ 
         141980 Dubna, Russia; E-mail: bedny@nusun.jinr.ru}
\author{H.V.~Klapdor-Kleingrothaus and E.Zaiti}
\address{Max-Planck-Institut f\"{u}r Kernphysik, \protect\\
        Postfach 103980, D-69029, Heidelberg, Germany \protect\\[3mm]}

\maketitle 

\begin{abstract} 
	Within the low-energy effective 
	Minimal Supersymmetric extension of Standard Model (effMSSM) 
	we calculated the neutralino relic density
 	taking into account slepton-neutralino and 
	neutralino-chargino/neutralino coannihilation channels. 
	We performed comparative study of these channels
	and obtained that both of them give 
	sizable contributions to the reduction of the relic density.
	Due to these coannihilation processes 
	some models (mostly with large neutralino masses)
	enter into the cosmologically interesting region 
	for relic density, but other models leave this region. 
	Nevertheless, in general, the predictions for 
	direct and indirect dark matter detection rates 
	are not strongly affected by
	these coannihilation channels in the effMSSM. 
\end{abstract} 

\section{Introduction} 
	Measurements of the cosmic microwave background radiation 
\cite{cmb} imply that the universe is almost flat
	$\Omega=\rho/\rho_c \simeq 1$, where  
	$\rho_c= 3 H^2/8\pi G_N$ is the critical closure density of 
	the universe, $G_N$ is Newton's constant and 
	$H=100\,h$ km/sec/Mpc is the 
	Hubble constant with $h = 0.7\pm 0.1$ 
\cite{wlf}. 
	A variety of data ranging from galactic rotation curves to 
	large scale structure formation and the cosmic microwave background
	radiation imply a significant density 
	$0.1< \oh <0.3$  
\cite{acc} of so-called cold dark matter (CDM). 
	It is generally believed that most of the CDM 
	is made of 
	weakly-interacting massive particles (WIMPs)
\cite{kt90}.  
	A commonly considered candidate for the WIMP 
	is the lightest neutralino, provided  
	it is the lightest supersymmetric particle (LSP) 
\cite{jkg}
	in the Minimal Supersymmetric extension of Standard Model (MSSM).  
	Four neutralinos in the MSSM being mass eigenstates 
	are mixtures of the 
	bino $\WT B$, wino $\WT W$ and higgsinos $\WT H_d^0$, $\WT H_u^0$,
	and the  LSP can be written as a composition
$\chi \equiv 
\WT \chi_1 = N_{11} \WT B + N_{12} \WT W + N_{13} \WT H_d^0 + N_{14} \WT H_u^0 
$,\
	where $N_{ij}$ are the entries of the neutralino mixing matrix. 
	In SUSY phenomenology one usually classifies 
	neutralinos as 
	gaugino-like (with $\pur \approx 1$),
	higgsino-like (with $\pur \approx 0$), and mixed, 
	where (gaugino) purity is defined as 
	$\pur = |N_{11}|^2 + |N_{12}|^2$.

	In most approaches the LSP is stable due to 
	R-parity conservation
\cite{susyreview}. 
	The neutralino, being massive, neutral and stable, 
	often provides a sizeable contribution to the relic density. 
	The contribution of neutralinos to the relic density is
	strongly  model-dependent and varies by several orders of magnitude 
	over the whole allowed parameter space of the MSSM.
	The neutralino 
	relic density then can impose stringent constraints on the
	parameters of the MSSM and the SUSY particle spectrum,
	and may have 
	important consequences both for studies of SUSY 
	at colliders and in astroparticle experiments
\cite{Belanger:2001fz}.
	In light of this and taking into account the 
	continuing improvements in determining the abundance 
	of CDM, 
	and other components of the Universe, which have now reached an
	unprecedented precision 
\cite{cmb},  
	one needs to be able to perform an accurate enough
	computation of the WIMP relic abundance, which would allow for a
	reliable comparison between theory and observation.

	In the early universe neutralinos existed in 
	thermal equilibrium with the cosmic thermal plasma.
	As the universe expanded and cooled, the thermal energy is 
	no longer sufficient to produce neutralinos at an appreciable rate, 
	they decouple and their number density 
	scales with co-moving volume. 
        The sparticles significantly heavier than the LSP
	decouple at the earlier time and
	decay into LSPs before the LSPs decouple themselves. 
	Nevertheless there may exist some other 
	next-to-lightest sparticles (NLSPs) 
	which are not much heavier than the stable LSP. 
	The number densities of
	the NLSPs have only slight Boltzmann suppressions with respect to the
	LSP number density when the LSP freezes out of chemical equilibrium
	with the thermal bath.
	Therefore they may still be present in the thermal plasma, 
	and NLSP-LSP and NLSP-NLSP interactions hold LSP in thermal
	equilibrium resulting with significant reduction of the LSP
	number density and leading to acceptable values even with a rather 
	heavy sparticle spectrum
\cite{EFOS-stau}.
	These coannihilation processes can be particularly 
	important when the LSP-LSP annihilation rate itself is suppressed. 

	The number density is governed by the Boltzmann equation
\cite{GondoloGelmini,EdsjoGondolo} 
\begin{equation} \label{boltzmann} 
{d n \over dt}+ 3 H n = - \sigmagen  (n^2 - n_{\eq}^2 ) 
\end{equation} 
      	with $n$ either being the LSP number density if there are no other 
      	coannihilating sparticles, or 
	the sum over the number densities of all coannihilation partners.
	The index ``eq'' denotes the corresponding equilibrium value.
	To solve the Boltzmann equation 
(\ref{boltzmann}) one needs
	to evaluate the thermally averaged neutralino 
	annihilation cross section $\sigmagen$. 
	Without coannihilation processes
	$\sigmagen$ is given as the thermal average of the LSP 
	annihilation cross-section 
	$\sigma_{\chi\chi}$ times relative velocity $v$ 
	of the annihilating LSPs
\begin{equation} \label{sigmano} 
\sigmagen = \sigmaav , 
\end{equation}
	otherwise, it is determined as
	$\sigmagen = \seff$, where  
	the effective thermally averaged cross-section
	is obtained by summation over coannihilating particles
\cite{GondoloGelmini,EdsjoGondolo}
\begin{equation} 
\seff = 
\sum_{ij} \sigmaijav {n^{\eq}_i \over n^{\eq} }{n^{\eq}_j \over n^{\eq}}.
\label{sigmaeff} 
\end{equation}
      	If $n_0$ denotes the 
nowadays
	number density of the relics, the relic density  is given by
\begin{equation} \label{finres} 
\Omega = {m_\chi n_0 \over \rho_{c}}.
\end{equation}

	Many increasingly sophisticated calculations of the relic density 
	of neutralinos in supersymmetric models, 
	with various approximations both in the evaluation of 
	$\seff$  and in solving the Boltzmann equation were performed
\cite{goldberg83,ehnos,krauss83,griest88,gkt,erl90,os91,DreesNojiri,%
	GriestSeckel,
	GondoloGelmini,an93,lny93,ows,%
	BaerBrhlik,%
	barb,%
	jkg,%
	Bottino,%
	leszek,%
	EdsjoGondolo,%
	Darksusy,%
	Ellis-Higgs,%
	EFOS-stau,%
	Belanger:2001fz,%
	Gomez:2000sj,%
	Lahanas:2000uy,%
	Arnowitt:2001yh,%
	Corsetti:2001yq,%
	Boehm:2000bj,%
	Ellis:2001nx,%
	Belanger:2001am,%
	Nihei:2001qs,%
	Baer:2002fv,%
	Nihei:2002ij}.
	Following 
\cite{Nihei:2002ij} 
	we briefly remind below of the major developments in the field.

	Perhaps for the first time strong constraints 
	for pure photino relic abundance were obtained in 
\cite{goldberg83}. 
	The first analysis of the general neutralino case was performed in  
\cite{ehnos,krauss83}. 
	Several other early papers subsequently appeared with more 
	detailed and elaborate analyses.  
	The two-neutralino annihilation into the ordinary fermion-antifermion 
	($f\bar f$) final states through the $Z$-exchange was computed 
	in detail
\cite{griest88}. 
	The first complete analysis of the
	neutralino annihilation into $W^+W^-$, $Z^0Z^0$ 
	and Higgs-pair final states was performed in
\cite{gkt}. 
	The Higgs contribution into neutralino annihilation
	was first computed in
\cite{gkt,erl90}.  
	For the pure gaugino-like and higgsino-like neutralinos
	(where several important resonances and final states are absent)
	all the annihilation channels were considered in 
\cite{os91}. 
	A first complete set of expressions for the 
	product of the cross section times velocity using the helicity 
	amplitude technique was computed in 
\cite{DreesNojiri}. 	 
	The relic density calculation was made by expanding the 
	$\langle \sigma v \rangle$ as a power series in neutralino velocity. 
	The angular and energy integrals in such a case 
	can be evaluated analytically and the 
	remaining integration over temperature was performed numerically.
	When expanded in the nonrelativistic limit, these give expressions 
	for the first two coefficients of the partial wave expansion.
	In the early papers the partial wave expansion of the
	$\langle \sigma v \rangle$ was used in most cases.  
	The method is normally expected to give an accurate enough 
	approximation in many regions 
	of the model parameter space because the relic neutralino 
	velocity is expected to be highly non-relativistic. 
	However, it fails near 	$s$-channel resonances 
(quite high-energetic)
	and thresholds for new final states,  
	as was first pointed out in 
\cite{GriestSeckel} and further emphasized in
\cite{GondoloGelmini,an93,lny93}. 
	In particular, it was shown
\cite{an93} that due to the very narrow width of the 
	lightest supersymmetric Higgs $h$, 
	in the vicinity of 
	its $s$-channel exchange the error can be as large as a few orders of
	magnitude.
	Therefore 
	a relativistic treatment of thermal averaging is 
	necessary. 
	A recent detailed analysis
\cite{Nihei:2001qs} showed that in the case of the often wide $s$-channel 
	resonance exchange of the pseudoscalar Higgs boson $A$, 
	the expansion produces a significant error
\cite{DreesNojiri}. 
	Furthermore subdominant channels and often neglected
	interference terms can also sometimes play a sizeable role. 

	The proper formalism for relativistic
	thermal averaging was developed in 
\cite{GondoloGelmini}, and used in 
\cite{BaerBrhlik}. 
	A more accurate treatment of the
	heat bath for both annihilating particles
	involving two separate thermal distributions was considered in 
\cite{GondoloGelmini,ows}. 
	A very useful compact expression for $\langle \sigma v \rangle$	
	as a single integral over the cross section
	was for the first time 	derived in 
\cite{GondoloGelmini}. 
	The DarkSusy code where the
	relic density of neutralinos is numerically computed without 
	the partial wave expansion approximation was developed in 
\cite{Darksusy}.

	An additional very strong reduction of the relic 
	abundance of WIMPs through coannihilation 
	was first discovered in 
\cite{GriestSeckel}. 
	There are regions in the MSSM parameter space where higgsino-like LSP, 
	light chargino and next-to-lightest neutralino 
	masses become nearly degenerate and
	all three species can exist in thermal equilibrium. 
	Their mutual coannihilation is often important, and even dominant 
\cite{GriestSeckel,Mizuta:1993qp,EdsjoGondolo}.
	In the coannihilation with the LSP can be involved any SUSY particle,
	provided its mass is almost degenerate with the mass of the LSP
\cite{GriestSeckel,Belanger:2001fz}.
	In the low-energy effective MSSM (effMSSM), 
	where one ignores restriction from unification assumptions and 
	investigates the MSSM parameter space at the weak scale
\cite{EdsjoGondolo,Bottino,BKKmodel}
	there is, in principle, no preference for the 
	next-to-lightest SUSY particle.
	Nevertheless due to quite reasonable and commonly used 
	sets of free parameters,
	when all gaugino mass eigenvalues are calculated in terms of 
	entries of gaugino mass matrices ($\mu, M_2, \tan\beta$), 
	the coannihilations between gauginos are expected 
	to occur most often, in the effMSSM
\cite{GriestSeckel,Mizuta:1993qp,EdsjoGondolo}.
	
	The relativistic thermal averaging formalism 
\cite{GondoloGelmini} was extended to include coannihilation processes in 
\cite{EdsjoGondolo}, and was implemented in the DarkSusy code
\cite{Darksusy} for coannihilation of charginos and heavier neutralinos.
	In was found 
\cite{EdsjoGondolo} that for higgsino-like LSP 
	such a coannihilation significantly decreases the relic density 
	and rules out these LSPs from the region of cosmological interest.
	For the highly bino-like LSPs, the reduction of the 
	relic density due to the coannihilation 
	is not strong enough to avoid an overclosing 
	of the universe.

  	The importance of the neutralino coannihilation 
	with sferminos was emphasized and investigated both for sleptons
\cite{EFOS-stau,Gomez:2000sj} or stops  
\cite{Boehm:2000bj,Ellis:2001nx} and sbottoms
\cite{Arnowitt:2001yh} in the so-called constrained MSSM (cMSSM) 
\cite{leszek,an93,Ellis:2002rp} or in supergravity (mSUGRA) models 
\cite{sugra}.

	In the most popular mSUGRA model 
\cite{sugra} SUSY breaking occurs in a hidden sector and is communicated 
	to observable sectors via gravitational interactions. 
	The model has a minimal set of parameters:
$
m_0,\ m_{1/2},\ A_0,\ \tan\beta\ {\rm and}\ {\rm sign}(\mu ) .
$
	Here $m_0$ is the universal scalar mass, 
	$m_{1/2}$ is the universal gaugino mass and 
	$A_0$ is the universal trilinear mass, all evaluated at $M_{\rm GUT}$, 
	$\tan\beta$ is the ratio of Higgs field vacuum expectation values
	and $\mu$ is a Higgs parameter of the superpotential.
 	In particular, within the framework of the mSUGRA, it was found
\cite{BaerBrhlik,Ellis-Higgs} that at large $\tan\beta$, 
	indeed large new regions of model parameter space gave 
	rise to reasonable values for the CDM relic density, 
	due to off-resonance neutralino
	annihilation through the broad $A$ and $H$ Higgs resonances.
	There are strong correlations of sfermion, 
	Higgs boson and gaugino masses in mSUGRA originating
	from unification assumptions. 
	In regions of mSUGRA parameter space where 
	$\chi$ and ${\tilde \tau}_1$ (or other sleptons) were nearly 
	degenerate (at low $m_0$), coannihilations could give 
	rise to reasonable values of the relic density even at 
	very large values of $m_{1/2}$, at both low and high $\tan\beta$
\cite{EFOS-stau,Arnowitt:2001yh}. 
	In addition, for large values of the parameter 
	$A_0$ or for non-universal scalar masses, 
	top or bottom squark masses could become nearly degenerate 
	with the $\chi$, so that squark coannihilation processes can become
	important as well
\cite{Boehm:2000bj,Ellis:2001nx}. 
	Therefore due to slepton and squark coannihilation effects, 
	the relic density can reach the cosmologically interesting range of 
	$0.1<\oh<0.3$ 
\cite{acc}.
	In the mSUGRA LSP is naturally almost pure bino-like as was 
	first noticed in 
\cite{chiasdm} from the point of view of low-energy SUSY and CDM.

\smallskip
	Having in mind investigation of future 
	prospects for direct and indirect detection of LSPs
	we follow the most phenomenological (general) view, 
	not bounded by theoretical restrictions from 
	sfermion/gaugino/Higgs mass unifications, etc. 
	To this end we need maximally general and accurate calculations 
	of the relic density within 
	the low-energy effective MSSM scheme (effMSSM)
\cite{Bottino,BKKmodel}.
	The only available high-level tools for these calculations 
	was the DarkSusy code (the best code to our
	knowledge at the moment, this paper was started).
	Unfortunately the code calculates only 
	neutralino with neutralino/chargino coannihilations (NCC), 
	which is not sufficient in the case of bino-like LSPs, when 
	neutralino-slepton coannihilation (SLC) and 
	neutralino-squark coannihilation are claimed to be dominant
\cite{EFOS-stau,%
	Gomez:2000sj,%
	Boehm:2000bj,%
	Ellis:2001nx,%
	Arnowitt:2001yh}.
	This paper is aimed at a 
	comparative study of NCC and SLC channels, 
	exploration of relevant changes in the relic density 
 	and investigation of their consequences for detection of  
	cold dark matter particles in the effMSSM. 

\section{The \lowercase{eff}MSSM approach}
	As free parameters in the effMSSM we use
\cite{BKKmodel}   	 
	the gaugino mass parameters $M_1, M_2$; 
	the entries to the squark and slepton mixing matrices 
	$m^2_{\t Q}, m^2_{\t U}, m^2_{\t D}, m^2_{\t R}, m^2_{\t L}$ 
	for the 1st and 2nd generations and 
	$m^2_{\t Q_3}, m^2_{\t T}, m^2_{\t B}, m^2_{\t R_3},m^2_{\t L_3}$
	for the 3rd generation, respectively; 
	the 3rd generation trilinear soft couplings $A_t , A_b , A_\tau$; 
	the mass $m_A$ of the pseudoscalar Higgs boson, 
	the Higgs superpotential parameter $\mu$, and $\tan\beta$.
	To reasonably reduce the parameter space we assumed 
$ m^2_{\t U} = m^2_{\t D} = m^2_{\t Q}$;
$ m^2_{\t T} = m^2_{\t B} = m^2_{\t Q_3}$;
$ m^2_{\t R} = m^2_{\t L}$;
$ m^2_{\t R_3} = m^2_{\t L_3}$
	and have fixed $A_b = A_{\t \tau} = 0$. 
	The remaining parameters defined our effMSSM 
	parameter space and were scanned randomly within the
	following intervals: 
\begin{eqnarray*} 
- 1 \tev < M_1 < 1\tev, \quad 
-2\tev < M_2 , \mu , A_t < 2\tev, \quad 
1.5<\tan\beta < 50, \\  
50\gev < M_A < 1000\gev, \quad
10 \gev^2 < m^2_{\t Q}, m^2_{\t L}, m^2_{\t Q_3}, m^2_{\t L_3} < 10^6\gev^2.
\end{eqnarray*}
	We have included the current experimental 
	upper limits on sparticle masses
	as given by the Particle Data Group 
\cite{pdg}. 
	As in 
\cite{EdsjoGondolo}, we have used the limit $\mchar \ge 85\gev$ 
	for the lighter chargino mass. The current limits on the rare 
	$b\rightarrow s \gamma$ decay 
\cite{flimits} and the $g-2$ limits 
\cite{glimits} have also been imposed. 
	In agreement with a flat accelerating universe 
\cite{acc}, we assume $0.1<\oh < 0.3$ 
	for the cosmologically interesting region.
	The calculations of the neutralino-nucleon cross sections, and direct 
	and indirect detection rates follow the description given in 
\cite{jkg,BKKmodel}.

	We have evaluated the relic density of the LSP under
	ignoring of any coannihilation (IGC), 
	taking into account only NCC or SLC separately, 
	as well as including both coannihilation channels (BCC). 
	To this end in our former code 
\cite{BKKmodel} DarkSusy procedures of $\seff$
	evaluation and solution of Boltzmann equation were implemented.
	All coannihilations with two-body 
	final states that occur between neutralinos, charginos and sleptons,
	as long as their masses are $m_i<2\mchi$ were included.
	The Feynman amplitudes for NCC and SLC were taken from DarkSusy 
\cite{Darksusy} and  
\cite{progfalk,EFOS-stau}, respectively. 
	We calculated $\seff$ and $\oh$ following 
	the relevant DarkSusy routines 
\cite{Darksusy}, which we have merged with code 
\cite{progfalk} in a way that guarantees the correct inclusion of SLC.

	If all sleptons, neutralinos and charginos in question 
	are substantially heavier than the LSP
	($m_i>2\mchi$) and no way for coannihilations,  
	the resulting relic density 
	$\Omega h^2 = \ohall = \ohds = \ohsl$ is equal to 
	$\Omega_\chi h^2$ of former results obtained 
	without any coannihilations (with $\seff = \sann$). 
	When at least one of 
	coannihilation channels (NCC or SLC) is indeed relevant, 
	the $\ohno$ (ignorance of any coannihilation) is calculated with 
\begin{equation} 
\seff_{\rm IGC} = \sann \left( n_\chi^\eq \over n^\eq \right)^2,
\end{equation}
	where $n^\eq$ includes {\em all}\ open coannihilation channels. 
	This formula allows a comparative study of results, relevant to one
	or both coannihilation channels, always delivering a 
	decreasing ratio $\ohco / \ohno < 1$ in accordance with 
\cite{Lahanas:2000uy} and sometimes contrary to 
\cite{EdsjoGondolo,EFOS-stau}. 
	We introduced $\ohco$ 
	as a common notation for $\ohall$,  $\ohds$ or $\ohsl$. 

\section{Results and Discussions}
	We performed our calculations in the effMSSM approach given above 
	and results of our considerations (scatter plots) are presented in  
Figs.~\ref{ratpap}--\ref{purpur2omb} and 
Figs.~\ref{namepap}--\ref{germratenam} for neutralino relic density 
	and CDM observables, respectively. 

\subsection{Relic density}
\begin{figure}[p] 
\begin{picture}(100,105)
\put(-12,-67){\includegraphics{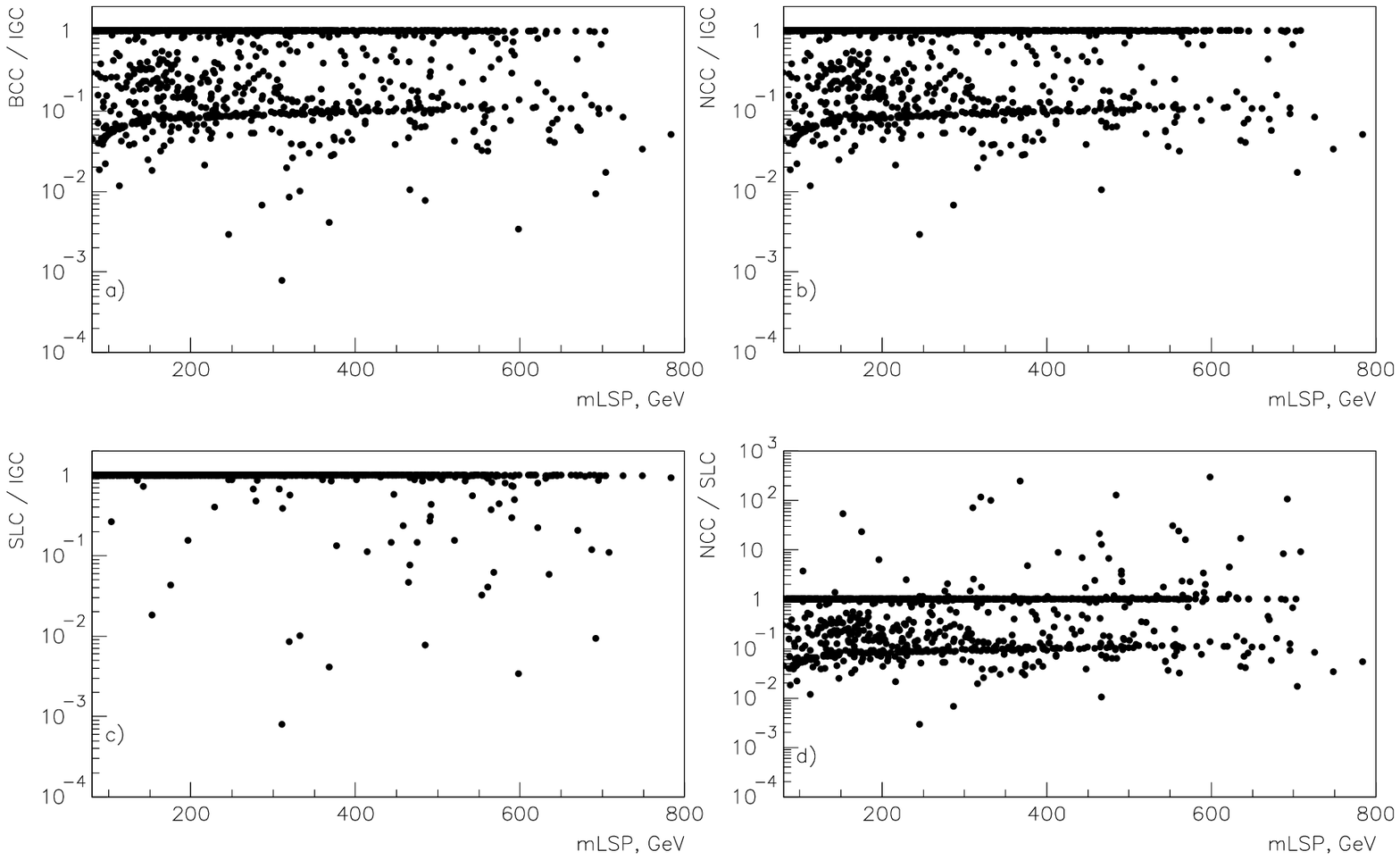}}
\end{picture}
\caption{Effects of neutralino-chargino/neutralino (NCC) and 
	slepton-neutralino (SLC) coannihilations. 
	Panels a)--d) display ratios 
	$\ohall / \ohno$, 
	$\ohds / \ohno$, 
	$\ohsl / \ohno$, and 
	$\ohds / \ohsl$, respectively.
	The maximal reduction factors for both NCC and SLC 
	are of the order of $10^{-3}$. 
\label{ratpap}}
\vfill 
\begin{picture}(100,100)
\put(-8,-65){\includegraphics{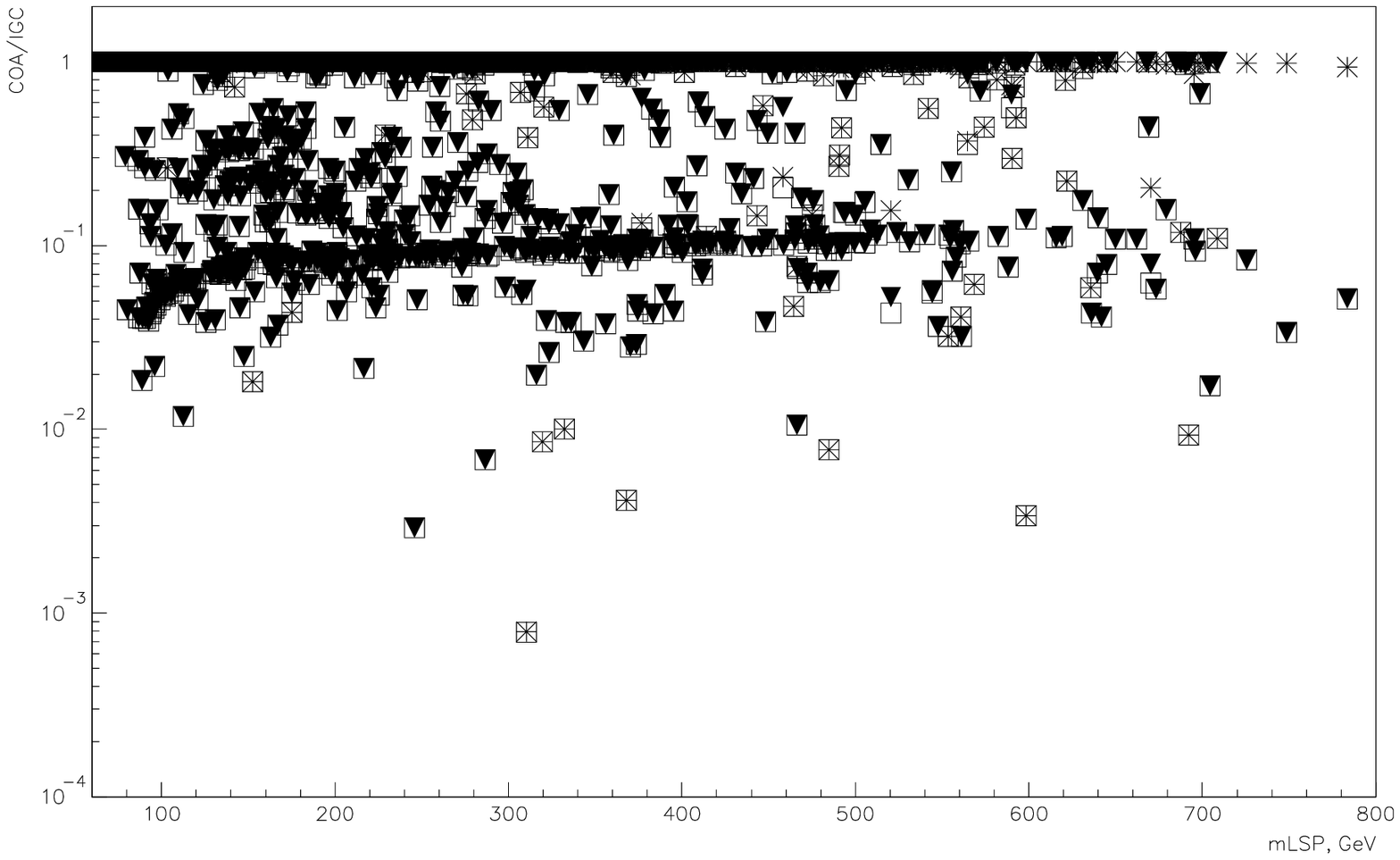}}
\end{picture}
\caption{The same as in 
	Fig.~\ref{ratpap} a), b) and c), but plotted together.
 	Here $\ohall / \ohno$, $\ohds / \ohno$, and $\ohsl / \ohno$ are marked 
	with squares, triangles, and stars, respectively. 
	Therefore, a square filled with a star (triangle) depicts 
	a model that is only affected by SLC (NCC), 
	while the other coannihilation channel 
	in the majority of models gives negligible contribution.
\label{rat43overomb}}
\end{figure}

	The general view of the reduction effect on the relic density (RD)
	due to NCC, SLC and BCC are shown in 
Fig.~\ref{ratpap} as ratios $\ohco / \ohno$
	together with comparison of NCC against  SLC in the form of 
	the ratio $\ohsl / \ohds$.  
	On the basis of our sampling (50000 models tested)
	the maximum factor of RD decrease due to NCC is about 
	$2\cdot 10^{-3}$ for $\mchi \approx 200\gev$, 
	while SLC reduces the RD maximally by a 
	factor of $8\cdot 10^{-4}$ for $\mchi \approx 300\gev$. 
	Both reduction factors are roughly of the same order of 
	about $10^{-3}$.
	These results depend on the applied experimental limits on the 
	second-lightest neutralino, chargino and stau masses.
	If there were no limits implemented on their masses, 
	the factor of relative RD reduction due to NCC  
	could reach a maximum value of $2\cdot 10^{-5}$ for 
	models with $\mchi \approx 40\gev$. 
	But in our case, the current experimental limits  are
	$\mchar > 85\gev$ and $\mstau > 81\gev$, 
	and therefore the critical LSP mass that enables 
	non-negligible NCC and SLC contributions 
	is also of the same order ($\mchi \ge 80\gev$).
	
\begin{figure}[h] 
\begin{picture}(100,90)
\put(-8,-70){\includegraphics{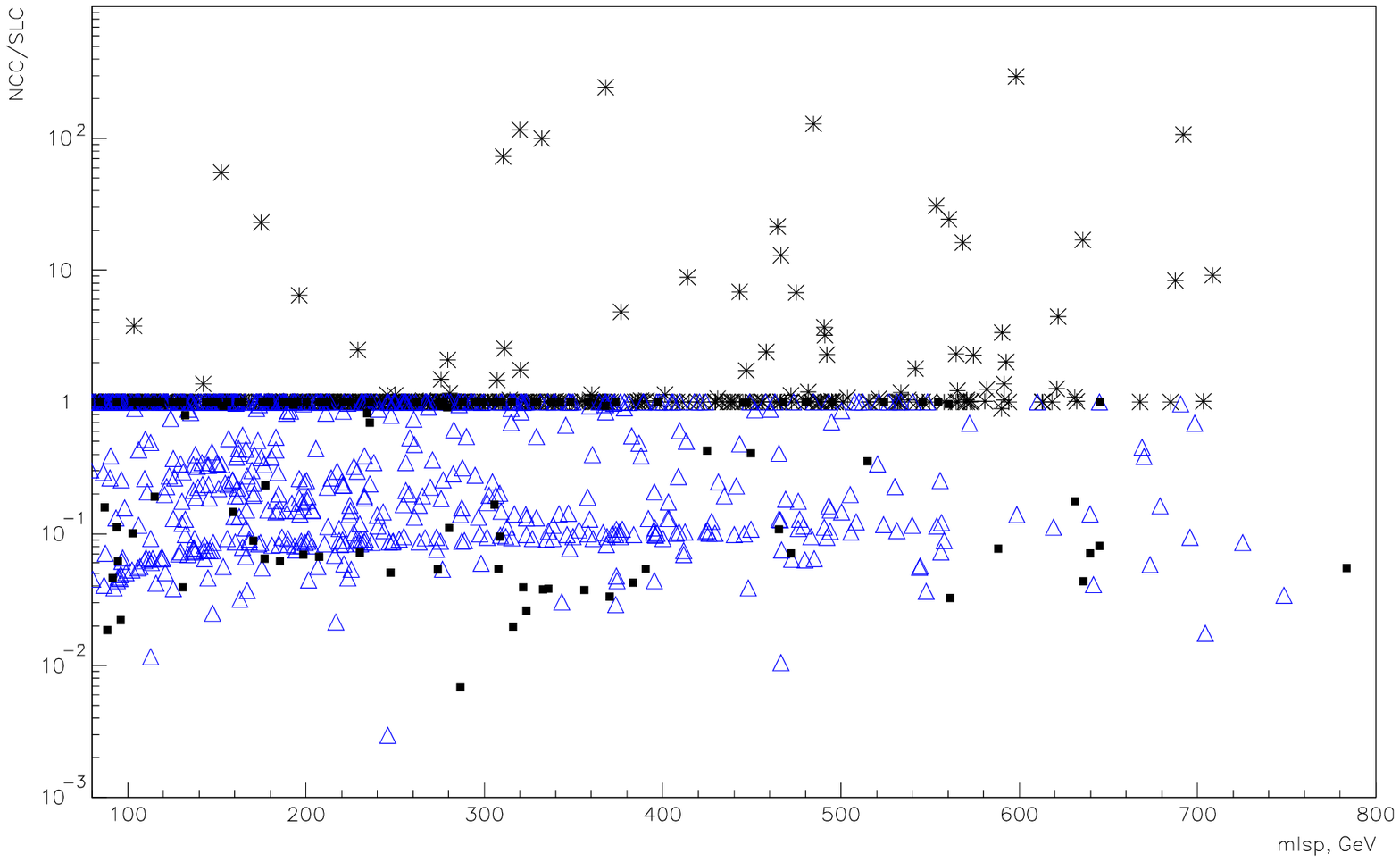}}
\end{picture}
\caption{Ratio $\ohds / \ohsl$ versus $\mchi$.
	Stars indicate that the $\t\tau$ is the NLSP, 
	triangles mean that the light chargino $\t\chi^\pm$  
	is the NLSP,  
	small filled squares mark the models where the 
	second-lightest neutralino $\t\chi_2$ is the NLSP. 
	One sees that if $\t\tau$ is the NLSP, 
	the SLC necessarily dominates, 
	while $\t\chi_2$ or $\t\chi^\pm$ being the NLSP 
	always leads to dominant NCC.
\label{sratncolnam}}
\bigskip
\begin{picture}(100,95)
\put(-8,-70){\includegraphics{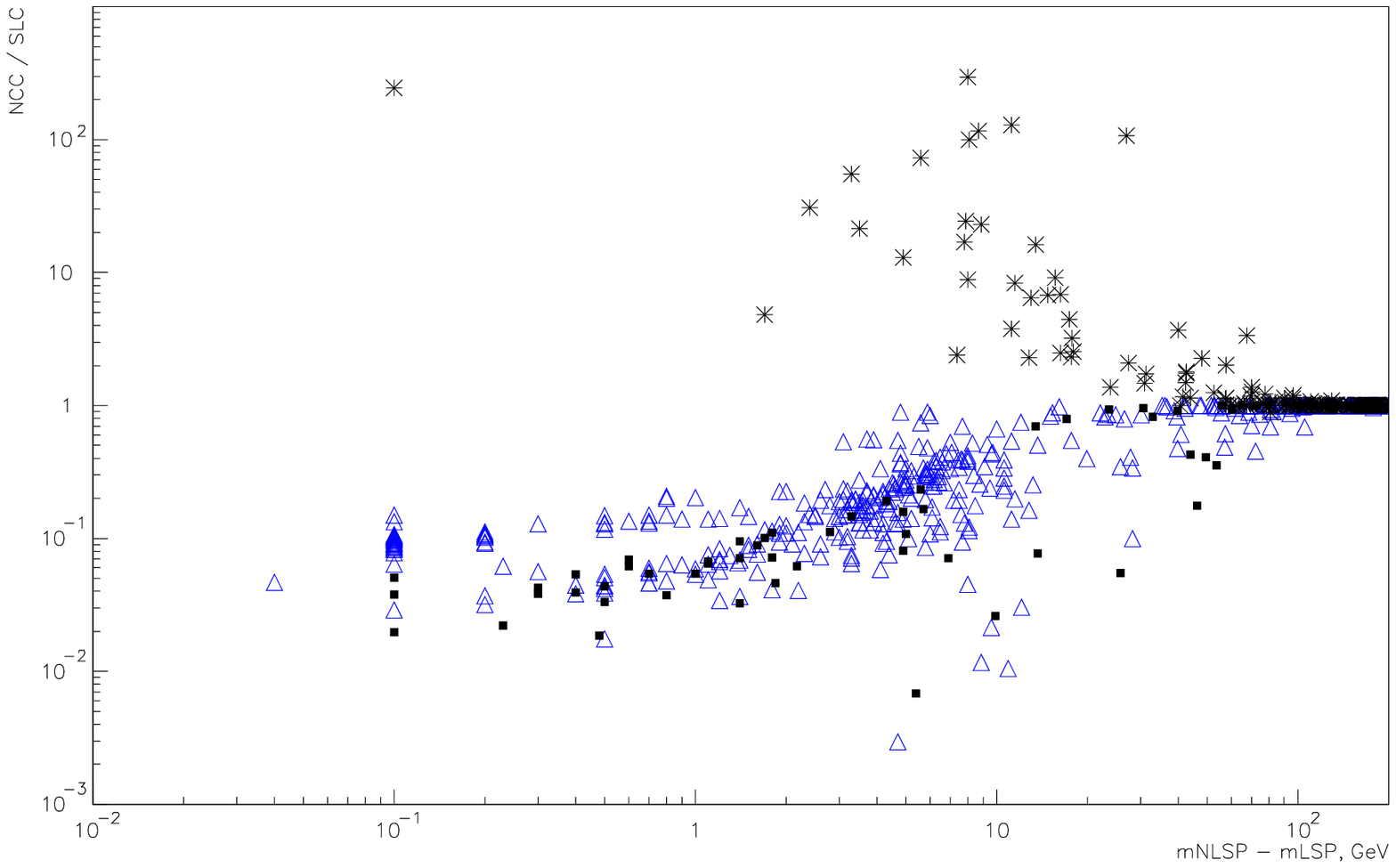}}
\end{picture}
\caption{The same as in 
Fig.~\ref{sratncolnam}, but versus $m^{}_{\rm NLSP} - \mchi$.
\label{fourncolnam}}
\end{figure}

	From 
Fig.~\ref{rat43overomb} one can see that the  
	reduction of RD by coannihilations 
	is mainly due to either NCC or SLC.
	The other channel of coannihilation plays no role or 
	leads only to a much smaller further reduction.  
	Although other coannihilation processes besides NLSP-LSP can 
	in principal occur 
	(including the next-to-NLSP (NNLSP) and next-to-NNLSP, etc), 
Fig.~\ref{sratncolnam} demonstrates that a stau $\t\tau$ as a
	NLSP indeed entails a dominant SLC effect, 
	while a next neutralino $\t\chi_2$ or 
	chargino $\t\chi^\pm$ as a  
	NLSP indeed entails a dominant NCC effect. 
 
	From 
Fig.~\ref{fourncolnam} one can notice that 
	mass differences $\mstau - \mchi < 20\gev$
	lead to a RD reduction factor of 0.5--0.005; 
	$20\gev < \mstau - \mchi < 40\gev$ 
	lead to the factor of 0.8--0.01 and mass differences 
	$ 40\gev<\mstau - \mchi<100\gev$ 
	can still lead to factors smaller than 0.3 due to SLC. 
	Mass differences 
	$m_{\t\chi_2,\t\chi^\pm} - \mchi 
	< 2\gev$ lead to a RD decrease 
	by factors of 0.1--0.005; 
	$2\gev<m_{\t\chi_2,\t\chi^\pm}-\mchi<30\gev$ 
	lead in general to factors 0.9--0.02 due to NCC. 
	For both kinds of NLSPs, the coannihilation effect may
	become negligible if $m_{\rm NLSP} - \mchi \ge 30\gev$, and
	necessarily becomes negligible if $m_{\rm NLSP} - \mchi \ge 100\gev$.
	Therefore, future increase of the 
	lower mass limits for all possible NLSP 
	(at Tevatron or LHC) can, in principle, strongly 
	reduce the importance of the effect of any of the 
	coannihilation channels. 

	Although we have implemented the 
	coannihilation opening threshold of $m_i = 2\mchi$, 	
	it was found that a SLC-reducing factor 
	less than 0.5 (0.1) occurs only for 
	$\mstau < 1.12\, (1.05)\, \mchi$. 
	Accordingly, a NCC-reducing factor less than 0.5 (0.1) appears for 
	$\mchar <1.16\, (1.10)\, \mchi$ and 
	$m_{\chi_2} < 1.11\, (1.08)\, \mchi$.
	Therefore for all channels of coannihilation, 
	relevant effects occur if the mass difference between 
	the coannihilation partner and the LSP is within 10--15\%. 
	This is in an agreement with previous considerations 
\cite{GriestSeckel,%
	EFOS-stau,%
	Lahanas:2000uy,%
	EdsjoGondolo,%
	Corsetti:2001yq,%
	Gomez:2000sj,%
	Ellis:2001nx,%
	Arnowitt:2001yh}.

	From
Fig.~\ref{fourncolnam} one can also see that 
	charginos and neutralinos come to lie close in mass to the 
	LSP more often than staus (and other sleptons).
	This is an expected result of 
	correlations in the gaugino sector of the effMSSM,
	as mentioned in the Introduction,  
	which explains the NCC dominance over SLC seen from 
Figs.~\ref{rat43overomb} and \ref{sratncolnam}.
	If one manages to construct a SUSY model where the
	LSP mass is almost always degenerate with one of the slepton masses  
	(as for example, in mSUGRA models with bino-like neutralinos)
	the dominant coannihilation channel will be SLC.

\begin{figure}[h] 
\begin{picture}(100,93)
\put(-8,-68){\includegraphics{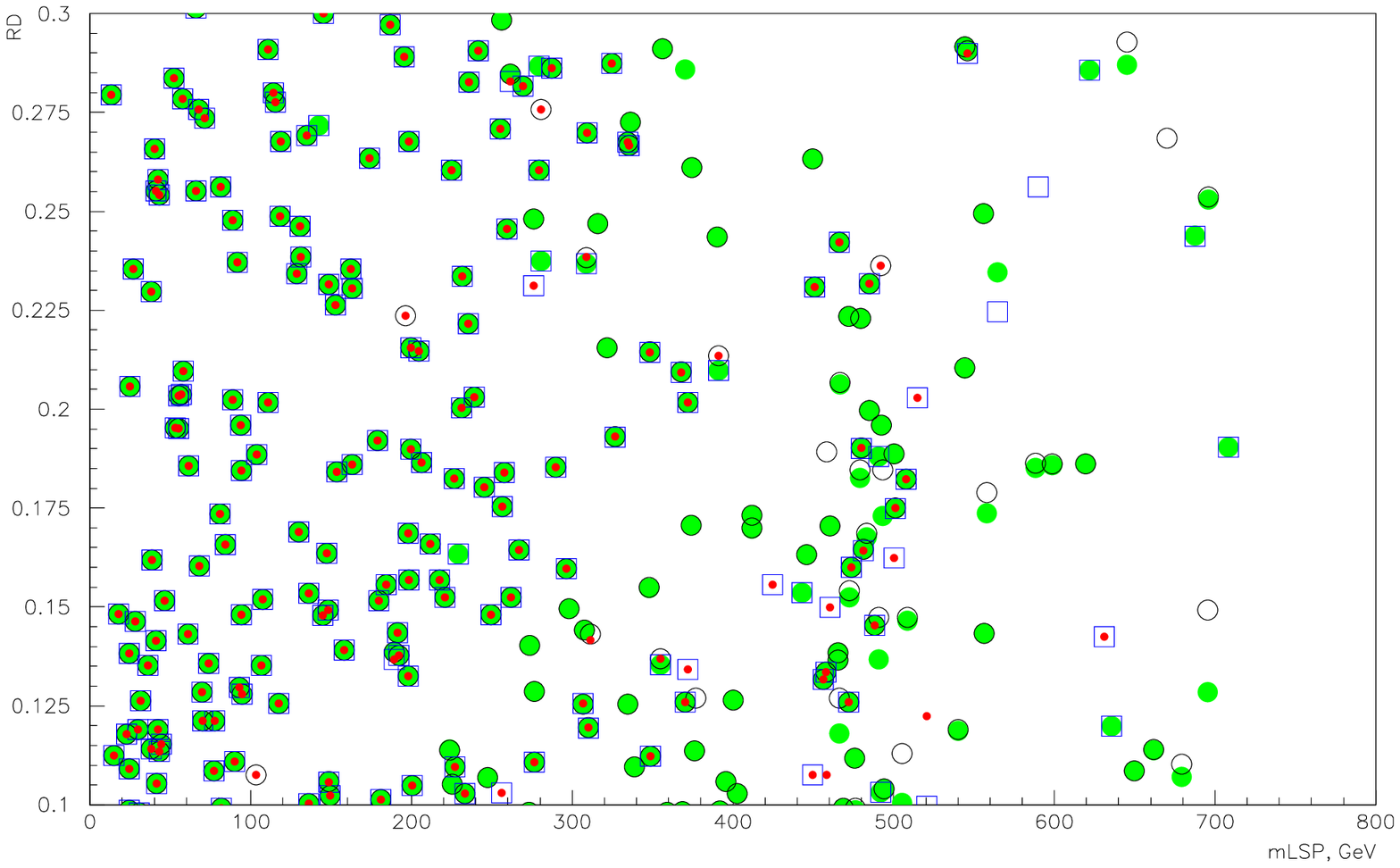}}
\end{picture}
\caption{Illustration of the shifting of effMSSM models inside 
	and outside the cosmologically interesting range 
	$0.1< \ohco < 0.3$ 
	due to NCC and SLC. 
	Models with $\ohno$, $\ohsl$, $\ohds$ and $\ohall$ are marked with 
	empty circles, filled circles, small dots, and empty squares, 
	respectively. 
	Therefore, a superposition of all those symbols corresponds 
	to a model which is totally untouched by coannihilation. 
	A black-framed filled circle marks a model which 
	is untouched by SLC ($\ohsl=\ohno$), but shifted down due to NCC. 
	If the corresponding $\ohall$ (which is equal to $\ohds$) 
	remains within this range, 
	it still presents in the figure 
	below the black-framed filled circle
	as an empty square with a black dot inside.
	By analogy, an empty square with a filled circle inside gives 
	a model which was shifted into the region due to SLC only
	($\ohall=\ohsl$), and if the corresponding $\ohno$ also is 
	in the cosmologically viable range, 
	it is located above the filled square 
	as an empty circle with a dot inside.
	One can notice 
	that a quite big amount of models is shifted out of 
	$0.1<\oh<0.3$ due to NCC (grey circles).
\label{ovfourgnam}}
\end{figure} 

\enlargethispage{0.8\baselineskip}
	In Fig.
\ref{ovfourgnam} all calculated 
	relic densities ($\ohno$, $\ohsl$, $\ohds$ and $\ohall$)
	are depicted in the cosmologically interesting region  
	$0.1< \ohco < 0.3$.
	There is a quite big amount of models (mostly with lower LSP masses)
	which are completely unaffected by coannihilation.
	When at least one of coannihilation channels is relevant, 
	the RD decreases and some cosmologically unviable models with  
	$\ohno > 0.3$ enter the cosmologically interesting range
	$0.1< \ohco < 0.3$,
	due to NCC (squares with a dot in the figure), 
	SLC (filled squares), 
	or both NCC and SLC (empty squares). 
	There are also models which enter the less interesting  
	region for LSP to be CDM ($\ohco < 0.1$).	
	The largest amount of models was shifted out 
	due to NCC (filled circles),
	and a relatively small amount of models was shifted out
	due to SLC (circles with dots), or both NCC and SLC (open circles).
	Contrary to mSUGRA
\cite{EFOS-stau}, in the effMSSM with SLC and NCC we can not find 
	a possibility to derive any cosmological upper limit for $\mchi$.
	There are cosmologically interesting LSPs 
	within the full mass range $12\gev < \mchi < 720\gev$
(Fig.~\ref{ovfourgnam}) accessible in our scan
	irrelevantly to neglecting or inclusion of any 
	coannihilation channels in question.

\begin{figure}[h] 
\begin{picture}(100,85)
\put(-4,-65){\includegraphics{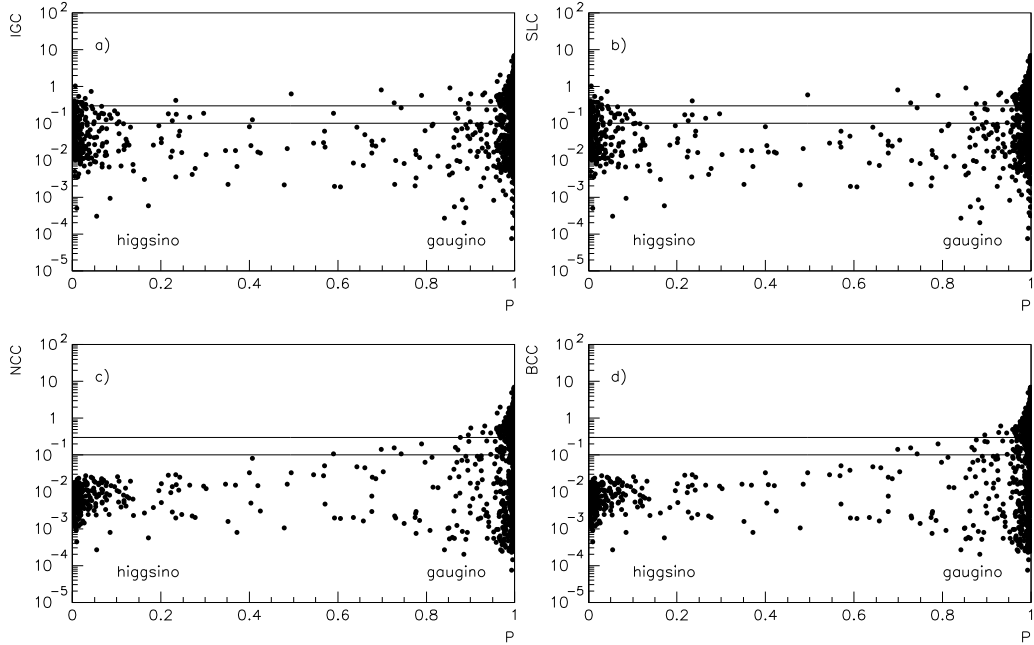}}
\end{picture}
\caption{Variation of relic density against gaugino purity
	as function of coannihilation channels included. 
	The NCC reduces the RD especially for 
	models with higgsino-like LSPs and shifts
  	these models out of cosmological interest. 
	The NCC together with the SLC leaves 
	only LSPs with  $\pur > 0.6$ 
	in the cosmologically interesting region.
\label{purpur2omb}}
\vspace*{-0.5\baselineskip}
\end{figure} 
	Cosmologically interesting LSPs occur with arbitrary compositions
	when coannihilations are ignored 
(Fig.~\ref{purpur2omb}), 
	the inclusion of NCC rules out
	all the models with higgsino-like LSPs, and SLC further tends to rule
	out LSPs with mixed composition, so that only LSPs with $\pur > 0.6$
	remain as dominant CDM candidates. 
	While NCC is important both for higgsino-like and gaugino-like LSPs 
	the SLC mainly affects only gaugino-like LSPs.
	In general our estimations 
(Fig.~\ref{purpur2omb})
	are in accordance with 
\cite{EFOS-stau} when bino-like LSPs and SLC are concerned 
	and in accordance with
\cite{EdsjoGondolo} when NCC effect is concerned.
\vspace*{-0.5\baselineskip}

\subsection{Detection rates}
\begin{figure}[h] 
\begin{picture}(100,92)
\put(-10,-68){\includegraphics{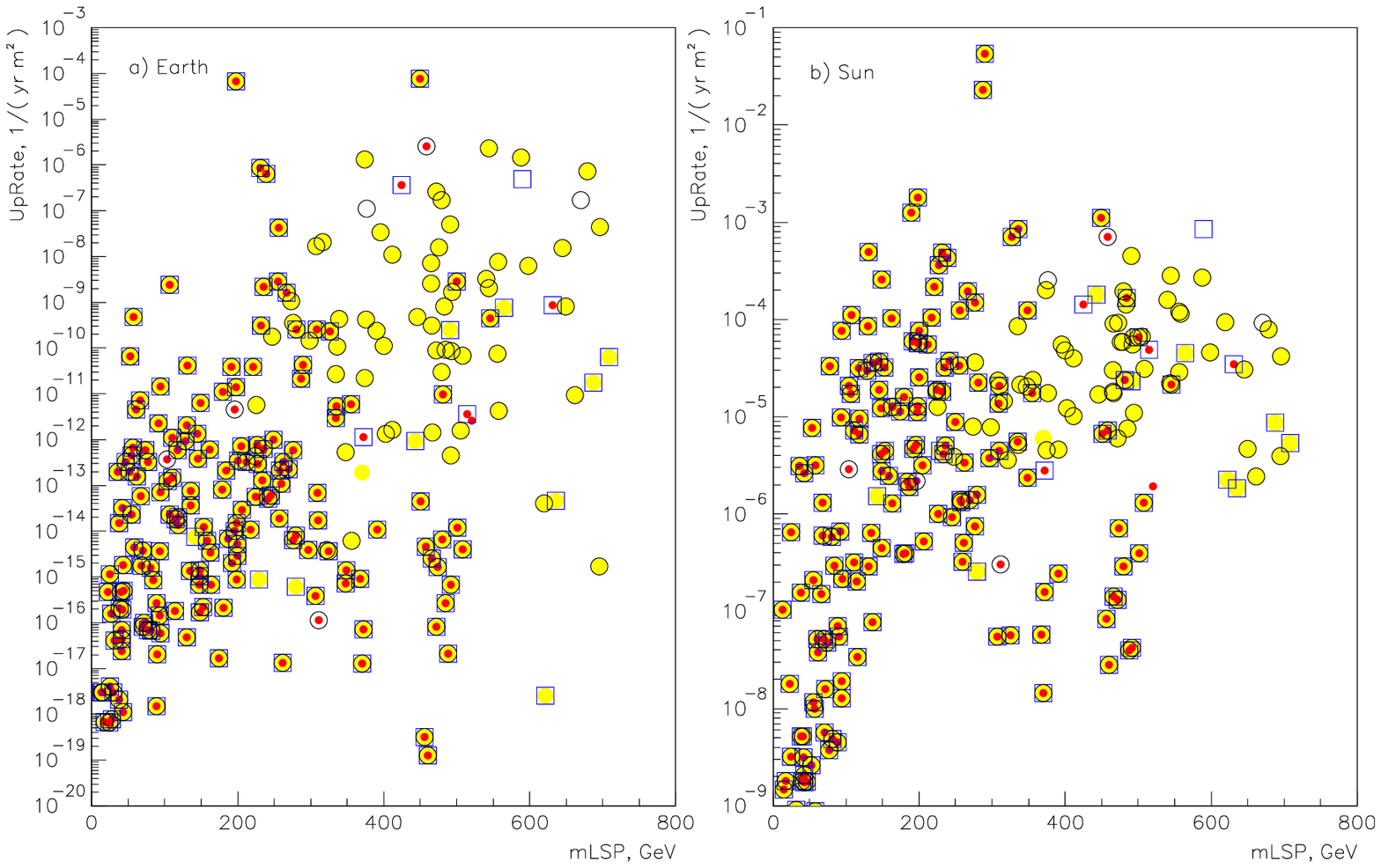}}
\end{picture}
\caption{Indirect detection rate for upgoing muons 
	from $\chi\chi$ annihilation in the Earth (a) and the Sun (b). 
	As in 
Fig.~\ref{ovfourgnam}, 
	empty circles, grey circles, black dots, and squares correspond to
	$0.1<\ohno, \ohsl, \ohds, \ohall<0.3$, respectively. 
	NCC slightly decreases the detection rates
	for models with $\mchi \ge 500\gev$.
	\label{namepap}}
\enlargethispage{0.7\baselineskip}
\vfill
\begin{picture}(100,94)
\put(-10,-70){\includegraphics{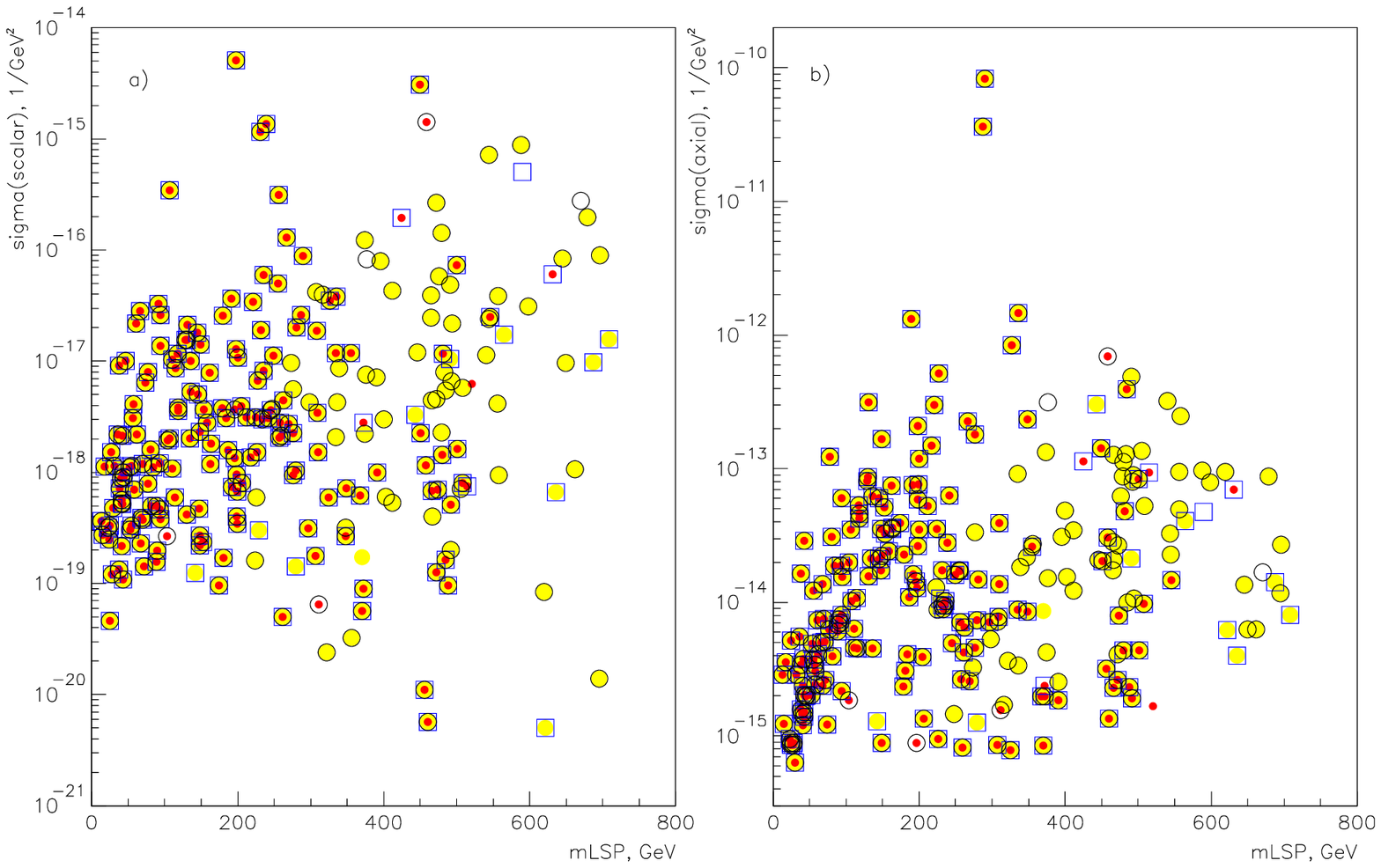}}
\end{picture}
\caption{Neutralino-proton scattering cross sections for 
	scalar (spin-independent) interaction
	(a) and axial (spin-dependent) interaction (b). 
	The notations as in 
Fig.~\ref{namepap}.	
\label{namwppap}}
\end{figure}

\enlargethispage{0.7\baselineskip}
	Now we briefly consider the influence of NCC and SLC on prospects for  
	indirect and direct CDM neutralino detection.
Figure~\ref{namepap} displays the expected indirect detection rates 
	for upgoing muons produced in the Earth by neutrinos 
	from decay products of $\chi\chi$ annihilation 
	which takes place in the core of the Earth or of the Sun.
	We compare the rate predictions 
	for cosmologically interesting LSPs when the RD 
	is evaluated with or without coannihilations taken into account. 
	We have seen before 
	that the RD in most models with 
	$\mchi \le 250\gev$ is untouched both by SLC and NCC, 
	because the difference  $m_{\rm NLSP} - \mchi$ 
	is too large to yield significant effects, 
	therefore
	the corresponding detection rates are not influenced
	(depicted in the figures as filled squares with dots inside). 
	For $\chi\chi$ annihilation in the Earth 
	upgoing muon detection rates merely lie within the range 
$ 10^{-19} \hgev < \Gamma^\mu < 5 \cdot 10^{-9}\hgev $ 
	as long as $\mchi \le 250\gev$. 
	When $\mchi \ge 250\gev$, some of the models with $0.1<\ohno < 0.3$ 
	are ruled out from the cosmological interesting range ($\ohco < 0.1$;
Fig.~\ref{ovfourgnam}) mainly due to NCC
(Fig.~\ref{namepap}).  
	Others models with $\ohno > 0.3$ are shifted inside this region
(Fig.~\ref{ovfourgnam} and 
 Fig.~\ref{namepap}) mainly due to SLC. 
	In total, for $\mchi \ge 250\gev$ one finds 
$ 10^{-19}\hgev < {\Gamma^\mu}_{\rm BCC} <5\cdot 10^{-7}\hgev $, 
	when the RD is evaluated with 
	coannihilations are taken into account and, 
	$10^{-19}\hgev < {\Gamma^\mu}_{\rm IGC} <4\cdot 10^{-6}\hgev$ 
	when coannihilations are neglected.
	The large values of the detection rates of
	$\chi\chi$ annihilation in the Earth are slightly decreased 
	(from $10^{-6}\hgev$ to $10^{-7}\hgev$) only for heavy LSPs 
	$\mchi > 500\gev$ 
	in accordance with the fact that  
	the corresponding models are ruled out from the cosmologically 
	interesting range. 
	The few models with maximal detection rates at 
	a level of $10^{-4}\,\hgev$ ($\mchi < 500\gev$)
	are found to be untouched. 

	In the case of indirect detection of 
	upgoing muons from $\chi\chi$ annihilation in the
	Sun one has in general a similar behavior  
	for models with $\mchi \ge 250-300\gev$.
	The only noticeable difference is  
	in the absolute predictions for 
	detection rates for models with 
	$\mchi > 600 \gev$ where instead of 
	${\Gamma^\mu}_{\rm IGC} < 10^{-4}\hgev$ 
	one expects the rates to be 
	${\Gamma^\mu}_{\rm BCC} < 3\cdot 10^{-5}\hgev$.
	The highest predicted detection rates of $10^{-1}\hgev$ are 
	again correlated to a few models 
	which are untouched by coannihilation.  

Figure~\ref{namwppap} shows neutralino-proton 
	scattering cross sections for the scalar 
	(spin-independent) and the axial (spin-dependent) 
	interactions.
	As in the previous figures 	
	the models with $\mchi \le 250\gev$ are hardly 
	affected by coannihilation, and 
	for the majority of those models 
	both neutralino-proton and neutralino-neutron scattering 
	cross sections reach values  
	$\sigma \le 10^{-17}\dtgev$ with the 
	maximal cross section of order $10^{-15}\dtgev$. 
	Cosmologically interesting models with 
	$\mchi \ge 250\gev$ were 
	influenced by coannihilations as discussed above, 
	and the maximal value of the neutralino-nucleon cross-section 
	decreases from $10^{-15}\dtgev$ to $5\cdot 10^{-16}\dtgev$
	for the models with $\mchi > 500\gev$. 
	In total, 
	independently of neglection or inclusion of NCC and SLC
	the maximal scalar scattering neutralino-nucleon
	cross section reaches 
	$10^{-16}$--$10^{-15}\dtgev$. 

	The spin-dependent neutralino-nucleon cross sections 
	are typically higher than the spin-independent ones, 
	and we have found the maximal values at 
$10^{-10}\dtgev$ for the axial neutralino-proton   and 
$10^{-11}\dtgev$ for the axial neutralino-neutron 
	scattering for the models 
	which are untouched by the coannihilations. 
	The majority of cosmologically interesting models yields 
	axial neutralino-proton scattering cross sections in the range 
$5\cdot 10^{-16}\dtgev < \sigma < 2\cdot 10^{-12}\dtgev$ 
	and  
	axial neutralino-neutron scattering cross sections in the range 
$2\cdot 10^{-16}\dtgev <\sigma <8\cdot 10^{-13}\dtgev $.

\begin{figure}[h] 
\begin{picture}(100,95)
\put(-10,-67){\includegraphics{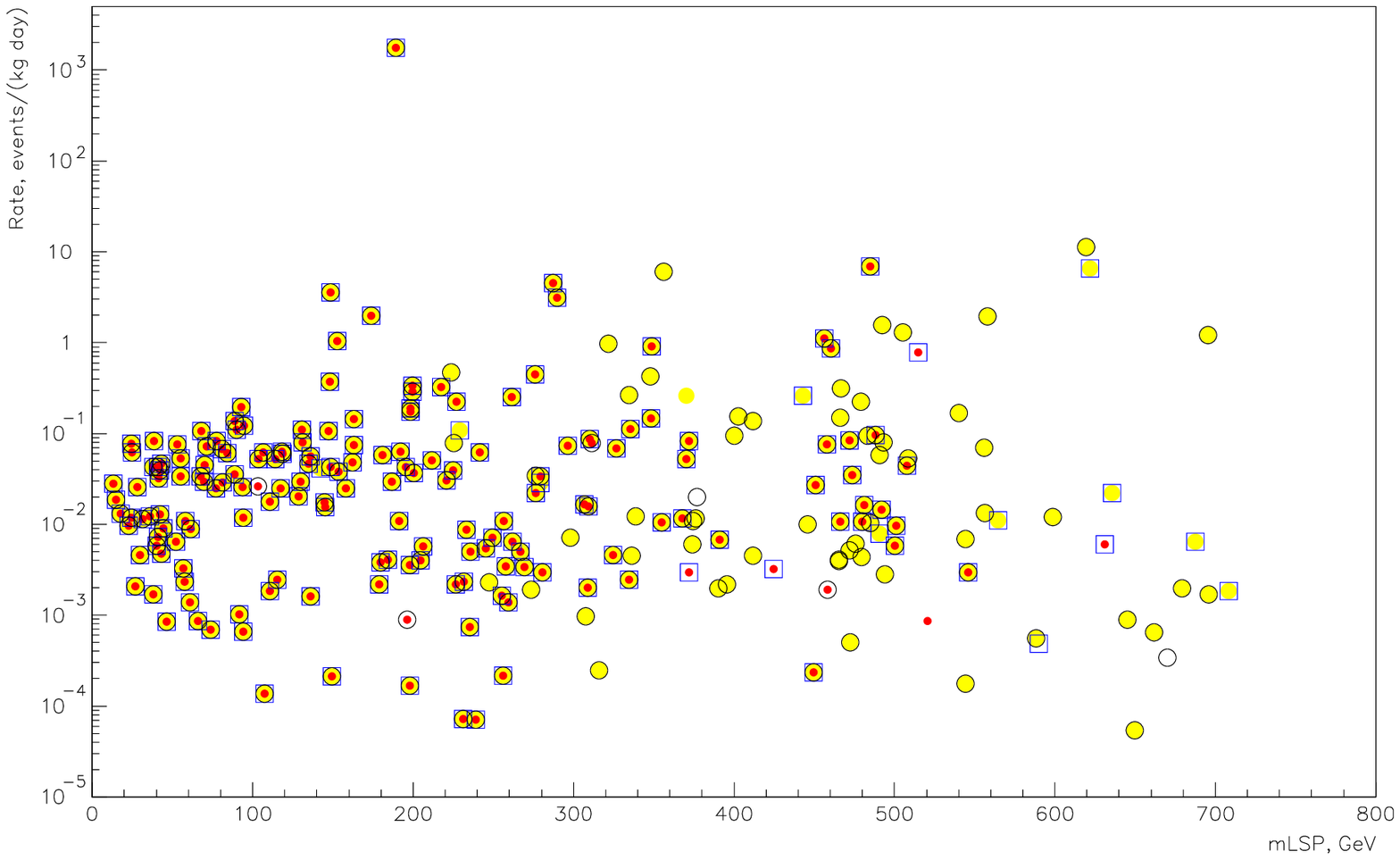}}
\end{picture}
\caption{Event rate for direct neutralino detection in a 
	${73\atop}{\rm Ge}$ detector. 
	As in 
Fig.~\ref{ovfourgnam}, 
	empty circles, grey circles, black dots, and squares correspond to
	$0.1<\ohno, \ohsl, \ohds, \ohall<0.3$, respectively. 
	NCC slightly decreases the maximal event 
	rates for models with $\mchi \ge 500\gev$, 
	but the models with smaller LSP mass 
	are untouched by the coannihilations.
\label{germratenam}}
\end{figure} 

Figure~\ref{germratenam} shows the expected direct detection event rates 
	calculated for a ${73\atop}{\rm Ge}$ detector
	when NSS, SLC, and BCC are taken into account. 
	For models with $\mchi \le 250\gev$ 
	coannihilations of any kind play no role.
	The optimistic estimations of the event rate for models with 
	$\mchi \ge 400\gev$ are slightly decreased due to NCC.

\section{Conclusions and Outlook} 
	Within the low-energy effective MSSM 
	we calculated the neutralino relic density (RD)
 	taking into account both slepton-neutralino (SLC) and 
	neutralino-chargino/neutralino (NCC) coannihilation channels. 
	To this end we have implemented in our code
\cite{BKKmodel} the
	relic density part (with neutralino-chargino coannihilations)
	of the DarkSusy code 
\cite{Darksusy} supplied with the adopted code of
\cite{EFOS-stau} (calculating slepton-neutralino coannohilations). 

	We have shown that in effMSSM the maximum factors of RD decrease 
	due to NCC as well as due to SLC can reach $10^{-3}$, 
	as long as the lower limits for 
	$\mstau$ and $\mchar$ are similar (both being of order of $80\gev$).
	We conclude that 
	both coannigilation effects are comparable in the effMSSM.
	For the majority of models affected by coannihilations and  
	successfully passed 
	all relevant accelerator, cosmological and rare-decay constraints
	it was observed that
	either NCC or SLC alone produces significant reduction of RD 
	while the other coannihilation channel gives  
	considerably smaller or zero reduction effect.
	Both coannihilations (NCC and/or SLC) 	
	are found to produce non-negligible effect only 
	if the relevant NLSP mass is smaller than $1.15\mchi$. 
	The type of the NLSP determines the dominant coannihilation channel. 
	Due to the fact that the effMSSM 
	more often favors neutralino and chargino, but not 
	sleptons to be the NLSP (the NLSP-LSP mass differences in general 
	are systematically larger for sleptons than for gauginos)
	the NCC is the more often dominant coannihilation channel
	in agreement with 
\cite{EdsjoGondolo}.
	Only LSPs with purity $\pur > 0.6$ 
	remain CDM candidates of cosmological interest.
	Some models with $\oh>0.3$ under neglection of coannihilation 
	enter into the cosmologically interesting region merely due to SLC, 
	and some other models 
	shift out of the region below $\oh<0.1$, merely due to NCC.
	In the effMSSM, contrary to mSUGRA 
\cite{EFOS-stau},
	both  coannihilations do not imply 
	new cosmological limits on the mass of the LSP.
	We noticed that
	the most optimistic predictions for neutralino-nucleon 
	cross sections, indirect and direct detection rates 
	for cosmologically interesting models 
	are untouched by these coannihilations. 
	Only for large $\mchi \ge 500\gev$, 
	the respectively high values are slightly reduced, 
	because of the NCC rules out corresponding 
	models from the cosmological interesting region 
	$0.1<\ohno<0.3$.

\smallskip 
	When our paper was almost finished we saw the preprints
\cite{Belanger:2001fz,Baer:2002fv,Nihei:2002ij}.
	In 
\cite{Belanger:2001fz} a new sophisticated code {\tt micrOMEGAs}
	for calculations of the relic density in the MSSM is presented.
 	The main characteristics of this code includes
	complete tree-level matrix elements for all subprocesses;
	{\em all}\ coannihilation channels with neutralinos, charginos, 
	sleptons, squarks and gluinos; 
	loop-corrected Higgs masses and widths.
	All calculations are performed with {\tt CompHEP}
\cite{comphep}. 
	In  
\cite{Baer:2002fv}
	the relic density of neutralinos in the mSUGRA 
	was calculated on the basis of annihilation diagrams, 
	involving sleptons, charginos, neutralinos and 
	third generation squarks. 
	The code {\tt CompHEP} 
\cite{comphep} was used, too, here.
	It was found in 
\cite{Baer:2002fv} 
	that coannihilation effects are only important on 
	the edges of the model parameter space, 
	where some amount of fine-tuning is necessary to
      	obtain a reasonable relic density. 
	This paper is mostly aimed at 
	prospects for SUSY search with 
	various $e^+e^-$ and hadron colliders 
	and pays no attention on the 
	interplay between different coannihilation channels.

In addition, 
	a full set of exact, analytic expressions for the 
	annihilation of the lightest neutralino pairs into all two-body
      	tree-level final states in the framework of minimal SUSY 
	is now available 
\cite{Nihei:2002ij}.
	This set of expressions does not rely on the partial wave
	expansion, includes all the terms and is valid 
	both near and further away 
	from resonances and thresholds for new final states.
	Further extension of this approach to 
	coannihilation processes together with the 
	above-mentioned {\tt CompHEP}-based codes
	will supply perhaps one of 
	most powerful tools for complete relic density calculations. 

\smallskip	
	V.B. thanks the Max Planck Institut fuer Kernphysik 
	for the hospitality and RFBR (Grant 00--02--17587) for support.
 


\begin{thebibliography}{999}\vspace*{-3.5\baselineskip}
\bibitem{cmb} 
	A.~T.~Lee {\it et al.} (MAXIMA Collab.),
	Astrophys.\ J.\  {\bf 561}, L1 (2001); 
	C.B. Netterfield {\it et al.} (BOOMERANG Collab.), astro-ph/0104460; 
	N.W. Halverson {\it et al.} (DASI Collab.), astro-ph/0104489; 
	P. de Bernardis {\it et al.}, astro-ph/0105296,
	A.~Melchiorri, astro-ph/0201237.
\bibitem{wlf} 
	W. L. Freedman, Phys. Rept. {\bf 333}, 13 (2000).
\bibitem{acc}
	{P. de Bernadis}, astro-ph/{0004404};
	{A. Balbi}, astro-ph/{0005124}.
\bibitem{kt90}
	E.W. Kolb and M.S. Turner, 
	{\em The Early Universe}, Addison-Wesley (1990);
	M.S. Turner, astro-ph/0108103.
\bibitem{jkg}  
	G.~Jungman, M.~Kamionkowski, and K.~Griest, 
	Phys.~Rep. {\bf 267} (1996) 195.
\bibitem{susyreview}
	H.P. Nilles, Phys.\ Rept. {\bf 110}, 1 (1984);
	H.E.~Haber and G.~Kane, Phys.\ Rept. {\bf 117}, 75 (1985);
	S. Martin, {hep-ph/9709356}. 
\bibitem{Belanger:2001fz} 
	G.~Belanger, F.~Boudjema, A.~Pukhov, and A.~Semenov,
	hep-ph/0112278.
\bibitem{EFOS-stau} 
	J.~R.~Ellis, T.~Falk, and K.~A.~Olive,
	Phys.\ Lett.\ B {\bf 444}, 367 (1998);
	J.~R.~Ellis, T.~Falk, K.~A.~Olive, and M.~Srednicki,
	Astropart.\ Phys.\  {\bf 13}, 181 (2000);
	[Erratum-ibid.\  {\bf 15}, 413 (2000)].
\bibitem{GondoloGelmini} 
	P. Gondolo and G. Gelmini, Nucl. Phys. B {\bf 360}, 145 (1991).
\bibitem{EdsjoGondolo} 
	J.~Edsjo and P.~Gondolo,
	Phys.\ Rev.\ D {\bf 56}, 1879 (1997);	
	Phys. Atom. Nucl. {\bf 61}, 1181, (1998);
	P.~Gondolo and J.~Edsjo,
	hep-ph/9804459;
	J.~Edsjo,
	hep-ph/9704384.
\bibitem{Darksusy} 
	({\tt DarkSUSY}).
	P. Gondolo, J. Edsj\"o, L. Bergstr\"om, P. Ullio, and E.A. Baltz
	astro-ph/0012234; 
	http://www.physto.se/~edsjo/darksusy/. 
\bibitem{goldberg83}
	H. Goldberg, Phys. Rev. Lett. {\bf 50}, 1419 (1983).
\bibitem{ehnos}
	J. Ellis, J. Hagelin, D. Nanopoulos, and M. Srednicki, 
	Phys. Lett. B {\bf 127}, 233 (1983);
	J. Ellis, J. Hagelin, D. Nanopoulos, K. Olive, and M. Srednicki, 
	Nucl. Phys. B {\bf 238}, 453 (1984).
\bibitem{krauss83}
	L.M.~Krauss, Nucl. Phys. B {\bf 227}, 556 (1983).
\bibitem{griest88}
	K. Griest, Phys. Rev. D {\bf 38}, {2357} ({1988});
	Erratum {\bf 39}, {3802} ({1989}).
\bibitem{gkt} 
	K. Griest, M. Kamionkowski, and M. Turner, 
	Phys. Rev. D {\bf 41}, 3565 (1990).
\bibitem{erl90}
	J. Ellis, L. Roszkowski, and Z. Lalak, 
	Phys. Lett. B {\bf 245}, {545} (1990).
\bibitem{os91}
	K.A. Olive and M. Srednicki, 
	Phys. Lett. B {\bf 230}, 78 (1989);
	Nucl. Phys. B {\bf 355}, 208 (1991).
\bibitem{DreesNojiri} 
	M. Drees and M. Nojiri, Phys. Rev. D {\bf 47}, 376 (1993).
\bibitem{GriestSeckel} 
	K. Griest and D. Seckel, Phys. Rev. D {\bf 43}, 3191 (1991).
\bibitem{an93} 
	P. Nath and R. Arnowitt, Phys. Rev. Lett. 
	{\bf 69}, 725  (1992); {\bf 70}, 3696 (1993);
	Phys. Lett. B {\bf 437}, 344 (1998).
\bibitem{lny93}
	J.L. Lopez, D.V. Nanopoulos, and K. Yuan, 
	Phys. Rev. D {\bf 48}, 2766 (1993).
\bibitem{ows} 
	M. Srednicki, R. Watkins, and K. Olive, 
	Nucl. Phys. B {\bf 310}, 693 (1988).
\bibitem{BaerBrhlik} 
	H. Baer and M. Brhlik, 
	Phys. Rev. D {\bf 53}, 597 (1996); {\bf 57}, 567 (1998); 
	H.~Baer \etal,  
	Phys.\ Rev.\ D {\bf 63}, 015007 (2001).
\bibitem{barb} 
	R. Barbieri, M. Frigeni, and G. F. Giudice, 
	Nucl. Phys. B {\bf 313}, 725 (1989).
\bibitem{Bottino} 
	A. Bottino, V. de Alfaro, N. Fornengo, G. Mignola, and S. Scopel, 
	Astropart. Phys. {\bf 1}, 61 (1992); 
	A. Bottino {\it et al.}, 
	Astropart. Phys. {\bf 2}, 67 (1994); 
	V. Berezinsky {\it et al.}, 
	Astropart. Phys. {\bf 5}, 1 (1996);
	A. Bottino, F. Donato, N. Fornengo, and S. Scopel, 
	Phys. Rev. D {\bf 59}, 095004 (1999).
\bibitem{leszek} 
	J. Ellis and L. Roszkowski, 
	Phys. Lett. B {\bf 283}, 252 (1992); 
	L. Roszkowski and R. Roberts, 
	Phys. Lett. B {\bf 309}, 329 (1993); 
	G. Kane, C. Kolda, L. Roszkowski, and J. Wells, 
	Phys. Rev. D {\bf 49}, 6173 (1994).
\bibitem{Ellis-Higgs} 
	J.~R.~Ellis, T.~Falk, G.~Ganis, K.~A.~Olive, and M.~Srednicki,
	Phys.\ Lett.\ B {\bf 510}, 236 (2001).
\bibitem{Gomez:2000sj}
	M.~E.~Gomez, G.~Lazarides, and C.~Pallis,
	Phys.\ Lett.\ B {\bf 487}, 313 (2000);
	Phys. Rev.  D {\bf 61}, 125312 (2000). 
\bibitem{Lahanas:2000uy} 
	A.~B.~Lahanas, D.~V.~Nanopoulos, and V.~C.~Spanos,
	Phys.\ Rev.\ D {\bf 62}, 023515 (2000).
\bibitem{Arnowitt:2001yh} 
	R.~Arnowitt, B.~Dutta, and Y.~Santoso,
	Nucl.\ Phys.\ B {\bf 606}, 59 (2001); 
	{R. Arnowitt, B. Dutta}, hep-ph/{0112157}.
\bibitem{Corsetti:2001yq} 
	A.~Corsetti and P.~Nath,
	Phys.\ Rev.\ D {\bf 64}, 125010 (2001); 
	A.~Corsetti and P.~Nath,
	hep-ph/0005234.
\bibitem{Boehm:2000bj} 
	C.~Boehm, A.~Djouadi, and M.~Drees,
	Phys.\ Rev.\ D {\bf 62}, 035012 (2000).
\bibitem{Ellis:2001nx} 
	J.~R.~Ellis, K.~A.~Olive, and Y.~Santoso,
	hep-ph/0112113.
\bibitem{Belanger:2001am}
	G.~Belanger, F.~Boudjema, A.~Cottrant, R.~M.~Godbole, and A.~Semenov,
	Phys.\ Lett.\ B {\bf 519}, 93 (2001). 
\bibitem{Nihei:2001qs} 
	T.~Nihei, L.~Roszkowski, and R.~Ruiz de Austri,
	JHEP {\bf 0105}, 063 (2001). 
\bibitem{Baer:2002fv}
	H.~Baer, C.~Balazs, and A.~Belyaev,
	hep-ph/0202076.
\bibitem{Nihei:2002ij} 
	T.~Nihei, L.~Roszkowski, and R.~Ruiz de Austri,
	hep-ph/0202009.
\bibitem{Mizuta:1993qp} 
	S.~Mizuta and M.~Yamaguchi,
	Phys.\ Lett.\ B {\bf 298}, 120 (1993). 
\bibitem{BKKmodel} 
	V.~A.~Bednyakov and H.~V.~Klapdor-Kleingrothaus,
	Phys.\ Rev.\ D {\bf 63}, 095005 (2001);
	Phys.\ Rev.\ D {\bf 62}, 043524 (2000);
	V.~A.~Bednyakov, H.~V.~Klapdor-Kleingrothaus, and H.~Tu,
	Phys.\ Rev.\ D {\bf 64}, 075004 (2001).
\bibitem{Ellis:2002rp}
	For current starus of CMSSM see
	J.~R.~Ellis, K.~Olive, and Y.~Santoso,
	hep-ph/0202110.
\bibitem{sugra} 
	A. Chamseddine, R. Arnowitt, and P. Nath,
	Phys. Rev. Lett. {\bf 49}, 970 (1982);
	R. Barbieri, S. Ferrara, and C. Savoy, 
	Phys. Lett. B {\bf 119}, 343 (1982);
	L.J. Hall, J. Lykken, and S. Weinberg, 
	Phys. Rev. {\bf D27}, 2359 (1983).
\bibitem{chiasdm} 
	L.~Roszkowski, Phys. Lett. B {\bf 262}, 59 (1991).
\bibitem{pdg} 
	D.~E.~Groom {\it et al.}  
	Eur.\ Phys.\ J.\ C {\bf 15}, 1 (2000).
\bibitem{flimits} 
	M. S. Alam \etal, (CLEO Collab.), 
	Phys.\ Rev.\ Lett.\ {\bf 74}, 2885 (1995); 
	K.~Abe {\it et al.},  (Belle Collab.), 
	hep-ex/0107065.
\bibitem{glimits}
	H.~N.~Brown {\it et al.}  (Muon g-2 Collab.),
	Phys.\ Rev.\ Lett.\  {\bf 86}, 2227 (2001).
\bibitem{progfalk} Toby Falk, private communication
\bibitem{comphep} 
	CompHEP. 
	A. Pukhov {\it et al.}, hep-ph/9908288;
	http://theory.sinp.msu.ru/\~{}pukhov/calchep.html.
\end{thebibliography}
\end{document}